\documentclass[aps,pra,twocolumn,groupedaddress,amsmath,amssymb]{revtex4}
\usepackage{graphicx}
\usepackage{bm}
\usepackage{multirow,amssymb,amsbsy,amsmath}

\newcommand{\Funktion}[2]{#1\kern-0.2em\left(#2\right)}

\begin{document}

\preprint{APS/123-QED}

\title{Nonlocal Hanbury Brown-Twiss Interferometry \& Entanglement Generation from Majorana Bound States}

\author{Sougato Bose$^{*}$ and Pasquale Sodano$^{\dagger}$}
\affiliation{
    ${}^{*}$Dept.~of Physics and Astronomy, %
    University College London, %
    Gower Street, %
    WC1E 6BT London, %
    United Kingdom,\\
    ${}^{\dagger}$Department of Physics and Sezione INFN,
    University of Perugia, Via A. Pascoli, 06123 Perugia, Italy,\\
    ${}^{\dagger}$Perimeter Institute of Theoretical Physics, Waterloo, Ontario N2L2Y5, Canada }
\author{}
\affiliation{
  }
\date{\today}

\begin{abstract}
We show that a one dimensional device supporting a pair of
Majorana bound states at its ends can produce remarkable
Hanbury Brown-Twiss like interference effects between well
separated Dirac fermions of pertinent energies. We find that the
simultaneous scattering of two incoming electrons or two incoming
holes from the Majorana bound states leads exclusively to an
electron-hole final state. This ``anti-bunching" in electron-hole
internal pseudospin space can be detected through current-current
correlations. Further, we show that, by scattering appropriate
spin polarized electrons from the Majorana bound states, one can
engineer a non-local entangler of electronic spins for quantum
information applications.  Both the above phenomena should be
observable in diverse physical systems enabling to detect the
presence of low energy Majorana modes.

\end{abstract}

 pacs{ 03.67.Bg, 07.60.Pb, 73.21.Hb, 74.78.Na}

\maketitle
\section{Introduction}

 Quantum Indistinguishability has striking
manifestations when two identical particles are brought together
at a beam splitter. For example, two bosons in identical states
would ``bunch" together when exiting a beam-splitter purely due to
interference effects \cite{Hong-Ou-Mandel}. Two fermions, on the
other hand, would exit separately or ``anti-bunch"
\cite{yamamoto}. These effects are indeed an instance of the
celebrated Hanbury Brown-Twiss effect, which has recently also
been tested with Helium atoms \cite{Aspect}. The same quantum
indistinguishability is exploited for the production of entangled
photons \cite{Shih-Alley}, and can also be used to entangle
generic massive particles \cite{Bose-Home}. Of course, all these
effects can occur only when the particles are brought together
spatially, for instance, at a beam splitter. It is thereby
interesting to look for settings where rather well separated
identical particles could manifest such phenomena.

 Here we report on the possibility of engineering a \emph{non-local} beam splitter enabling the above class of phenomena for distant charged
 fermions. Here, by
 ``non-local" we mean spatially extended. Going beyond the usual two particle interference in orbital/momentum space, here one finds a
 {\em Hanbury Brown-Twiss effect in the
electron-hole internal
 pseudospin space}. This is enabled by Majorana mid-gap low energy modes which transform between electrons and holes \cite{Fu-Kane2},
 effectively making them indistinguishable in a scattering experiment. This Hanbury Brown-Twiss effect is thereby a detector of the Majorana modes.

Recently, low energy Majorana (neutral charge self-conjugated
fermion) modes located at the edges of linear devices have been
predicted to induce non-local phenomena
\cite{Sodano,Beenakker,Fu}. Indeed there are a variety of
platforms to realize such devices such as a quantum wire immersed
in a p-wave superconductor \cite{Sodano,Kitaev2}, cold-atomic
systems mimicking p-wave superconductors \cite{Das-Sarma1},
topological insulator-superconductor-magnet structures
\cite{Beenakker,Fu-Kane1,Refael2} and potentially also
semiconductor systems \cite{Das-Sarma2,Refael1}. The evidences of
their non-local nature are distance independent tunneling
\cite{Sodano}, crossed Andreev reflection \cite{Beenakker} and
teleportation-like coherent transfer of a fermion \cite{Fu}.
Finally, they may be easily manipulated \cite{Fu-Kane2} and are
relevant excitations also in conventional superconductors
\cite{JackiwPi}. So far, the primary application envisaged for
these fermions has been topological quantum computation
\cite{Kitaev1}. As the second key result of this paper we will
show another use of these modes, namely that Majorana bound-states
(MBS) could be used to engineer entanglement between the spins of
well separated particles, a pivotal resource in quantum
information.

The paper is organized as follows. In Section II we consider the
scattering of Dirac fermions off the edge MBS of energy $E_M$ in
the spinless model investigated in \cite{Beenakker}. Here, we show
that, when the energy of the incoming fermions is nearly {\em resonant}
with $E_M$, the edge MBS induce a beam splitting process which acts like
 an equally weighted 4-port beam splitter, with ports corresponding to
 both spatial and electron-hole isospin states. In Section III we show that with two incident Dirac fermions, this allow
for {\em fermion antibunching} in the pseudospin space which has holes
and electrons as its two states. This is one of the central results of our paper. In Section IV we determine the
signature of this fermion antibunching in the zero-frequency
spectral density of the current fluctuations in the leads. Section
V generalizes the results of section II by accounting for the spins
of the fermions. In section VI, we show that the edge MBS allow for
generating entanglement between the spins of distant electrons
only by pertinently choosing the polarizations of the incoming
fermions. In Section VII we analyze a few condensed matter settings
where our findings may be helpful in detecting the presence of
MBS; section VIII is devoted to a few remarks on our results.
For a reader interested more in the logical steps leading to our 
  central results, rather than the full technical details, we recommend
  focussing primarily on sections III and VI, taking the relevant scattering matrices from the sections immediately preceding them.

\begin{figure}
\centering \includegraphics[width=0.35 \textwidth]{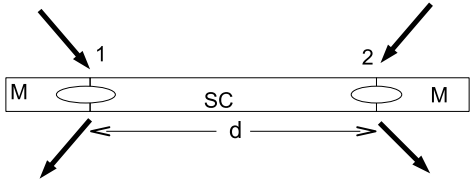}
\caption{Non-local beam splitter and electron spin entangler. The
MBS are shown as empty ellipses $1$ and $2$. One specific
realization where MBS occur at boundaries between magnets (M) and
superconductors (SC) deposited on quantum spin-Hall insulators is
depicted, though our results hold more generally. Incoming and
outgoing particles are shown by arrows, and may, in practice, be
tunneled in/out by STM tips or electron pumps acting as leads.}
\label{Fig1}
\end{figure}

\section{Nearly resonant electron (hole) scattering from Majorana edge
states}

We consider a one dimensional device supporting two weakly coupled
MBS at its ends as shown in Fig.\ref{Fig1}. The MBS are labeled as
$1$ and $2$ and schematically shown as empty ellipses in the
figure. As the separation between the MBS increases their energy $
E_M$ decreases exponentially \cite{Sodano}. For the sake of
clarity, we will first show how this device produces
Hanbury Brown-Twiss like interference effects between spatially
separated Dirac fermions in the {\em spinless} models investigated
in \cite{Kitaev2,Sodano}; later, we show how all the results are
valid for more realistic spinfull physical settings
\cite{Fu-Kane1,Das-Sarma2,Refael1}. The Hamiltonian describing the
weak coupling between the MBS $\gamma_1$ and $\gamma_2$ is given
by
\begin{equation}
H_M=i E_M \gamma_1\gamma_2,
\end{equation}
where $\gamma_j$ are Majorana operators defined by
$\gamma_j=\gamma_j^{\dagger}$ and satisfying
$\gamma_j\gamma_k+\gamma_k\gamma_j=2\delta_{kj}$ 
(in our definition, the $\gamma_j=c_j+c_j^{\dagger}$ in terms of 
Dirac fermionic operators $c_j$).

Leads, also labeled as $1$ and $2$, are connected to the device as
shown in Fig.\ref{Fig1}, allowing for the scattering of Dirac
fermions (electrons or holes) from each of the MBS; we further
assume that lead $1$ is coupled only to bound state $1$ while lead
$2$ only to bound state $2$. The Hamiltonian describing the leads
needs not to be specified at this stage since- for the evaluation
of the S-matrix- the leads my be effectively removed by
introducing complex embedding potentials \cite{datta}. Using the
approach developed in \cite{datta,ghur} one may then write the
unitary scattering matrix $S$ in a form which is formally
independent on the model used to describe the leads; for the
scattering of fermions from the MBS located at the the ends of the
quantum wire shown in Fig.\ref{Fig1} the $S$ matrix has been
computed in \cite{Beenakker} as
\begin{equation}
S(E)=1+2\pi i W^{\dagger}(H_M-E-i\pi WW^{\dagger})^{-1}W
\end{equation}
where $W$ is a rectangular matrix 
\begin{equation}
\left( \begin{array}{cccc}
~w_1 & 0 & w_1^{\star} & 0  \\
~0 & ~w_2 & ~0 & w_2^{\star} 
\end{array} \right)
\nonumber
\end{equation}
in the basis $\{|e_1\rangle,|e_2\rangle,|h_1\rangle,|h_2\rangle\}$, with
$|e_j\rangle$ and $|h_j\rangle$ representing an electron and a
hole in the lead $j$. $W$
describes the coupling of the scatterer ($H_M$) to the
leads and $E$ is the energy of the incident electrons/holes. The
entries $w_j$ of the $W$ matrix are related to the
couplings to the leads $\Gamma_j=2\pi w_j^2$ \cite{Beenakker}.

 For our purposes, it is convenient to assume that $E>>\Gamma_j$, as well
as $E\approx E_M$ (i.e., the energies of the incoming Dirac fermions are tuned to be nearly resonant with the Majorana coupling energy). Under
these circumstances, the $S$ matrix simplifies to
\begin{equation} S=\frac{1}{2}\left( \begin{array}{cccc}
~1 & -i & -1 & -i  \\
~i & ~1 & ~i & -1 \\
 -1 & -i & ~1 & -i \\
~i & -1 & ~i & ~1
\end{array} \right)\label{SMat}\end{equation}
where the basis is, again,
$\{|e_1\rangle,|e_2\rangle,|h_1\rangle,|h_2\rangle\}$. Note that this regime is {\em different}
from the one considered by Akhmerov et. al. \cite{Beenakker},
where only the terms corresponding to crossed Andreev reflection
(i.e $\langle h_2| S |e_1\rangle$ and $\langle h_1| S
|e_2\rangle$) are maximized. Here, we work in a regime where all
the entries of $S$ have the same magnitude. It is the {\em implications} of this scattering matrix $S$ of Eq.(\ref{SMat}) that we work out in this paper.
It is this $S$ that enables both the non-local Hanbury-Brown Twiss interferometry in isospin space and the non-local entanglement generation.

 Let us first illustrate the action of the above $S$ matrix by describing what happens to electrons tunneling in from one end of the one dimensional device. If, at time $t$, a single electron tunnels into the Majorana mode located at site 1, i.e., the incoming state  is $c^{\dagger}_1|0\rangle$,
 it transforms, under $S$, to
 \begin{equation}
 c^{\dagger}_1|0\rangle \xrightarrow{\text{MBS}} \frac{1}{2}(c^{\dagger}_1+i c^{\dagger}_2-d^{\dagger}_1+id^{\dagger}_2)|0\rangle,
 \label{mbsscat}
 \end{equation}
 where $c^{\dagger}_j$ ($d_j^{\dagger}$) creates an electron (hole) at site $j$. In Eq.(\ref{mbsscat}), we have used MBS above the arrow to indicate
 that Majorana bound states are
 responsible for the process. Since the transformation (\ref{mbsscat}) is equivalent to a four port beam splitter, with MBS inducing the beam
 splitting process, one can equally
 well take MBS to stand for ``Majorana Beam-Splitter". Eq.(\ref{mbsscat}) implies that an incoming electron has $\frac{1}{4}$th probability of coming
 out of each site as an electron or a hole. If another electron scatters at a different time $t'$ on the Majorana mode located at position 2, it will
 also scatter with exactly the same probabilities for the four possible outcomes. The joint probability for two incoming electrons to exit as two
 electrons or two holes (whichever the output port) would thus be $\frac{1}{2}$.

 \section{Hanbury-Brown effect in pseudo-spin space}
 We will now show that when
 $t=t'$, i.e., simultaneous scattering, two particle interference can take place so that the probability of two electrons or two holes exiting
 is completely suppressed. By $t=t'$ we mean that the wavepackets of the two incoming electrons (holes) are large enough so
 that their time of arrival cannot be distinguished when one observes them after the scattering.

 When two electrons scatter {\em simultaneously}, one at site 1 and the other at site 2, one has
 \begin{eqnarray}
 c^{\dagger}_1c^{\dagger}_2|0\rangle &\xrightarrow{MBS}& \frac{1}{2}(c^{\dagger}_1+i c^{\dagger}_2-d^{\dagger}_1 \nonumber\\ &&+ id^{\dagger}_2)~~\frac{1}{2}(-ic^{\dagger}_1+ c^{\dagger}_2-i d^{\dagger}_1-d^{\dagger}_2)|0\rangle \nonumber\\&=& \frac{1}{2} (i c^{\dagger}_1 d^{\dagger}_1 - c^{\dagger}_1 d^{\dagger}_2 + c^{\dagger}_2 d^{\dagger}_1 +i c^{\dagger}_2 d^{\dagger}_2)|0\rangle.
 \label{mbsscat2}
 \end{eqnarray}
In the last step of Eq.(\ref{mbsscat2}), we have used
$d^{\dagger}_j(E)=c_j(-E)$ (which effectively embodies the
indistinguishability between an electron and a hole), where $E$ is
energy. From Eq.(\ref{mbsscat2}) one sees that the probability for
two outgoing electrons (holes) after the scattering is zero.
Exactly the {\em same} holds when two holes scatter simultaneously
at leads 1 and 2. This is an interference effect in the same sense
as the anti-bunching of fermions at a normal two port beam
splitter, where fermions cannot exit through the same port.
Instead of being in the spatial channels, here the anti-bunching
is in the internal pseudospin space which has particle and hole as
its two states. The unitary conversion of an electron to a hole,
is, {\em per se}, not surprising in view of Refs.\cite{Fu-Kane2}.

 Of course, in a practical realization, the condition $E\sim E_M$ required for obtaining the scattering matrix $S$ of Eq.(\ref{SMat}) may not be
 exactly met. To see the effect of
 an energy mismatch, we denote by $\delta E$ the amount by which  $E$ deviates (either positively or negatively) from $E_M$; this deviation is,
 however, assumed to be much lower
 than $E_M$ itself (i.e., $\delta E<< E_M$). Without assuming $\delta E<< E_M$, one may end up in qualitatively different regimes: e.g.,
 for $\delta E$ comparable to $-E_M$, one reaches the regime of Ref.\cite{Beenakker} of only crossed
 transmission.
 For $\delta E<< E_M$, the scattering matrix as a function of $\delta E$ is given by
 \begin{equation} S_{\delta E}= \frac{i\Gamma}{\delta E+i\Gamma} S + \frac{\delta E}{\delta E+i\Gamma} I
\label{SdelE}\end{equation} where $\Gamma=\Gamma_1\sim \Gamma_2$
and $I$ is the $4\times 4$ identity matrix. In deriving
Eq.(\ref{SdelE}), one ignores the second and higher powers of both
$\delta E/E_M$ and $\Gamma/E_M$ as $E_M>>\Delta E, \Gamma$. It is
easy to check that, despite the above approximation, $S_{\delta
E}$ is unitary; furthermore, Eq.(\ref{SdelE}) holds for any value
of the ratio $\delta E/\Gamma$ as long as $E_M>>\Delta E, \Gamma$.
Using $S_{\delta E}$, one readily obtains that the probability of
observing an electron-electron output state becomes finite and
equal to $\frac{(\delta E)^2}{(\delta E)^2+\Gamma^2}$, which, of
course, vanishes when $E\sim E_M$.

Before ending this section, as a brief aside, we point out that, when one electron and one hole scatter at sites $1$ and $2$ respectively, for $\delta E<<\Gamma$, the
  incoming state $
 c^{\dagger}_1d^{\dagger}_2|0\rangle$ evolves to $\frac{1}{2} (- c^{\dagger}_1 c^{\dagger}_2 - i c^{\dagger}_1 d^{\dagger}_1 + i c^{\dagger}_2 d^{\dagger}_2 - d^{\dagger}_1 d^{\dagger}_2)|0\rangle$,
 implying the interferometric vanishing of the probability of one outgoing electron and one outgoing hole in separate
 leads. We will not delve further into this case, but next proceed to discuss the signatures of the Hanbury-Brown interferometry in the case of two incident electrons.

\section{Spectral density of current fluctuations: a signature of fermion
antibunching in pseudospin space}

In the previous section we have described the scattering as a
process where one sends particles one by one through the leads at
specific times. However, in practice, rather than controlling
times, one could control the energies $\epsilon_1$ and
$\epsilon_2$ of the particles in their respective leads, so as to
make them behave indistinguishably when
$\epsilon_1\sim\epsilon_2$. Then, the standard way to observe the
predicted fermion anti-bunching is through a measure of the
correlations between the currents in leads $1$ and $2$. The
current in lead $j$ may be written as \cite{buttiker,loss-burkard}
\begin{equation}
I_j(t)=\frac{e}{h\nu}\sum_{\epsilon,\epsilon^{'}} e^{i(\epsilon-\epsilon^{'})t}\{a_j^{\dagger}(\epsilon)a_j(\epsilon^{'})-b_j^{\dagger}(\epsilon)b_j(\epsilon^{'})\}
\end{equation}
where $a_j$ and $b_j$ denote the incoming and outgoing particles
and $\epsilon$ and $\epsilon^{'}$ are the energies of the
particles and $\nu$ is the density of states of the incoming
electrons. The spectral density of the current fluctuations
$\delta I_j=I_j-\langle I_j \rangle$ between the leads at zero
frequency is \cite{loss-burkard}
\begin{equation}
P_{ij}=\text{lim}_{~T\rightarrow \infty} \frac{h\nu}{T} \int_0^T dt~ \text{Re} \langle \delta I_1 (t) \delta I_2 (0)\rangle.
\end{equation}
Using $S_{\delta E}$ of Eq.(\ref{SdelE}) and considering an
incoming two electron state
$c^{\dagger}_1(\epsilon_1)c^{\dagger}_2(\epsilon_2)|0\rangle$,
where $c^{\dagger}_j(\epsilon_j)$ denotes an electron of energy
$\epsilon_j$ in lead $j$, one finds
\begin{equation}
P_{ij}=\frac{e^2}{h\nu}\frac{\Gamma^2}{\{(\delta E)^2+\Gamma^2\}^2}\{(\delta E)^2-\Gamma^2\}\delta_{\epsilon_1,\epsilon_2},
\end{equation}
where $\delta_{\epsilon_1,\epsilon_2}$ is the Kronecker delta
function. Note that, when the incident electrons are
distinguishable i.e., $\epsilon_1\neq\epsilon_2$, then, as
expected, $P_{ij}=0$ since for an electron exiting one lead there
could equally well be an electron or a hole exiting the other
lead. When, instead, $\epsilon_1=\epsilon_2$ (i.e., the particles
are indistinguishable), then for $|\delta E|<|\Gamma|$, the
domination of the electron-hole final state (as in
Eq.(\ref{mbsscat2})) makes $P_{ij}<0$, which allows to the detect
the predicted ``anti-bunching" in pseudospin space. For $|\delta
E|>|\Gamma|$ a process of amplitude $\Gamma\delta E$ in which only
one of the electrons scatter, while the other remains in its lead,
dominates; Fermi statistics now makes the electrons anti-bunch
spatially (the more conventional antibunching
\cite{yamamoto,loss-burkard}), contributing to a positive
$P_{ij}$. As in Ref.\cite{Beenakker}, our results are not
inconsistent with those of Bolech and Demler \cite{Bolech-Demler},
since their results apply when the energy of the incoming
electrons is much higher than $E_M$.

\section{Scattering Matrix in the spinfull case}

So far our analysis has been confined to the spinless model
investigated in \cite{Sodano, Beenakker}, while for the promising
implementations \cite{Fu-Kane1,Das-Sarma2,Refael2,Refael1}, the
Majorana modes should involve superpositions of operators of
different spins. For example, for a realization in a
ferromagnet-s-wave superconductor-ferromagnet structure on a
quantum spin-Hall edge \cite{Refael2}, one has
\begin{eqnarray}
\gamma_1=\frac{1}{\sqrt{2}}(c_{1,\uparrow}-i c_{1,\downarrow}+ic_{1,\downarrow}^{\dagger}+c_{1,\downarrow}^{\dagger})\nonumber\\
\gamma_2=\frac{1}{\sqrt{2}}(c_{2,\uparrow}+i
c_{2,\downarrow}-ic_{2,\downarrow}^{\dagger}+c_{2,\downarrow}^{\dagger}),
\label{spinfull}
\end{eqnarray}
where $c_{j,\sigma}$ creates an electron with spin $\sigma$ in lead $j$. Defining the spin states
$|\pm y\rangle=\frac{1}{\sqrt{2}}(|\uparrow\rangle\pm i|\downarrow\rangle)$, and using the
basis $\{|e_{1,+y}\rangle,|e_{2,-y}\rangle,|h_{1,+y}\rangle,|h_{2,-y}\rangle,|e_{1,-y}\rangle,
|e_{2,+y}\rangle,|h_{1,-y}\rangle,$
 $|h_{2,+y}\rangle\}$,
the scattering matrix is found to be
\begin{equation}
S_{\text{spinfull}}=\left( \begin{array}{cc}
~I & ~{\bf 0}  \\
~{\bf 0} & ~S
\end{array} \right),
\label{scatspinfull}
\end{equation}
where in (\ref{scatspinfull}), $I$ and ${\bf 0}$ are the $4\times 4$ Identity and null matrices, while $S$ is the scattering matrix given by Eq.(\ref{SMat}).

 When one uses $S_{\text{spinfull}}$ to study the scattering of the incident state $c_{1,-y}^{\dagger}c_{2,+y}^{\dagger}|0\rangle$, one only needs
 the lower-right $4\times 4$ block of $S_{\text{spinfull}}$. Thus, precisely the same electron-hole output state as in Eq.(\ref{mbsscat2})
 is obtained, apart from the fact that, now, the spin indices $-y$ and $+y$ are pinned to the sites $1$ and $2$ respectively. Thus, by choosing the
 spin polarizations of the incoming electrons pertinently, one can observe {\em all} the effects described till now. This should be possible in
 a variety of systems as Majorana modes of the form given by Eq.(\ref{spinfull}) are quite generic, e.g., also realizable in
 semiconductor-superconductor-magnet structures \cite{Refael1}.

\section{Entanglement of distant electron spins}

 We now propose a protocol for the generation of entanglement between spins of well
 separated particles incoming at site 1 and at site 2. For this purpose, we choose the realization of Majorana fermions given by Eq.(\ref{spinfull})
 and make two electrons with
 parallel spins in the $\uparrow$ direction come in {\em simultaneously} i.e., choose the initial
 state $c_{1,\uparrow}^{\dagger}c_{2,\uparrow}^{\dagger}|0\rangle$. Then, using $S_{\text{spinfull}}$,
 one gets
 \begin{eqnarray}
 c_{1,\uparrow}^{\dagger}c_{2,\uparrow}^{\dagger}|0\rangle &\xrightarrow{MBS}&  \frac{1}{4} (c_{1,\uparrow}^{\dagger}c_{2,\uparrow}^{\dagger}-c_{1,\downarrow}^{\dagger}c_{2,\downarrow}^{\dagger}+2 c_{1,\uparrow}^{\dagger}c_{2,\downarrow}^{\dagger}\nonumber\\&+& ...)|0\rangle,
 \label{ent}
 \end{eqnarray}
where $...$ denotes terms such as
$c_{1,\sigma}^{\dagger}c_{1,\sigma^{'}}^{\dagger},
c_{2,\sigma}^{\dagger}c_{2,\sigma^{'}}^{\dagger},
c_{j,\sigma}^{\dagger}d_{k,\sigma^{'}}^{\dagger}$ and
$d_{j,\sigma}^{\dagger}d_{k,\sigma^{'}}^{\dagger}$, which are not
relevant to our discussion. Eq.(\ref{ent}) implies that, when two
outgoing electrons are obtained in leads $1$ and $2$ separately,
their state is
$|\xi\rangle_{12}=\frac{1}{\sqrt{6}}(|\uparrow\rangle_1|\uparrow\rangle_2-|\downarrow\rangle_1|\downarrow\rangle_2+2
|\uparrow\rangle_1|\downarrow\rangle_2)$ where, as it is usually
done \cite{Bose-Home,loss-burkard}, one uses the lead label to
label the electron. $|\xi\rangle_{12}$ is an entangled state of
the spins of electrons $1$ and $2$, with the amount of
entanglement (as quantified by the von Neumann entropy of one of
the particles \cite{entconc}) being $0.19$ ebits. Though the
entanglement is not very high, $|\xi\rangle_{12}$ is a pure state,
and hence of value in quantum information, as its entanglement can
be concentrated without loss by local means \cite{entconc}.
Moreover, the probability of obtaining two outgoing electrons in
separate leads (i.e., $|\xi\rangle_{12}$) is rather high, namely
$3/8$. At the expense of decreasing this probability, one may
improve the degree of entanglement of the generated state by
tuning the polarizations of the incoming electrons. For instance,
if the incoming state is
$(\frac{1}{\sqrt{10}}c_{1,+y}^{\dagger}+\frac{3}{\sqrt{10}}c_{1,-y}^{\dagger})(\frac{3}{\sqrt{10}}c_{2,+y}^{\dagger}+\frac{1}{\sqrt{10}}c_{2,-y}^{\dagger})|0\rangle$,
one obtains an output state of entanglement $0.75$ ebits, while
the probability of the generation this state becomes $0.055$. The
spin entanglement of the outgoing electrons could be measured by
passing them through separate spin filters as in Ref.
\cite{martin}.

Unlike the entanglement generation scheme of Ref.
\cite{Bose-Home}, here particles polarized parallel to each other
suffice to generate entanglement. Importantly, in our protocol,
particles at a distance from each other can be made entangled;
this may avoid the decoherence arising necessarily from the
transport needed to separate the particles after a local
entangling mechanism. In addition, the distance between the
entangled particles can be enhanced by putting $n$ copies of our
setup in series with leads connecting the end of one copy to the
beginning of another. The probability of obtaining the state
$|\xi\rangle_{12}$ in the leftmost and rightmost leads will then
be $(3/8)^n$.

 One might think that in analogy with Ref.\cite{loss-burkard}, perhaps it is 
possible also to detect entanglement between distant electronic
spins by injecting to the opposite ends of our one dimensional device. As a maximally entangled state of two spins can be written in any basis, let us consider the $|\pm y\rangle$ basis for spins.
In this basis, two of the maximally entangled states can be written as $|\psi^{\pm}\rangle_{12}=(c^{\dagger}_{1,+y}c^{\dagger}_{2,-y}\pm c^{\dagger}_{1,-y}c^{\dagger}_{2,+y})|0\rangle$. The detection of entanglement at a normal 50-50 beam splitter relies crucially on {\em both} the incoming states $c^{\dagger}_{1,+y}c^{\dagger}_{2,-y}|0\rangle$ and $c^{\dagger}_{1,-y}c^{\dagger}_{2,+y}|0\rangle$ evolving at the beam-splitter and interference ({\em i.e.}, cancellation/addition) between the terms resulting from the evolution each of the above two states. Only as a result of these cancellations/additions does the bunching/anti-bunching effects  evidencing entanglement arise. However, here the term $c^{\dagger}_{1,+y}c^{\dagger}_{2,-y}|0\rangle$ does not even evolve under the action of $S_{\text{spinfull}}$ so that interferences are impossible. Thus though our device can generate spin entanglement it cannot detect spin entanglement.

\section{A few condensed matter settings}

One simple setting where the non-local two particle interferometry
and the entanglement generation between distant electrons from MBS
may be observed can be engineered with
magnet-superconductor-magnet junctions deposited on the edge of a
2D quantum spin Hall insulator \cite{Beenakker,Refael2}. Just as
in Ref.\cite{Beenakker}, one can observe these effects when the
Majorana modes are separated by a distance $d$ of several micrometers at
temperatures of the order of 10 mK. For this setting, the explicit
form of the Majorana operators is exactly the same as in
Eq.(\ref{spinfull}) \cite{Refael2}. Interestingly, strong
spin-orbit coupled quantum wires in proximity with ferromagnets
and superconductors also support the realization of MBS
\cite{Das-Sarma2,Refael1} given in Eq.(\ref{spinfull})
\cite{Refael1}. As in previous proposals
\cite{Bolech-Demler,Das-Sarma2,Refael2}, also in these settings,
two STM tips could act as the leads $1$ and $2$ to observe the
non-local two particle Hanbury Brown-Twiss interferometry. For the
entanglement generation, instead, it will be more useful to have
synchronized electron pumps \cite{Pepper} feeding in the incoming
electrons. In addition, the filtering of the desired state
$|\xi\rangle_{12}$ can be achieved by pumps capturing exactly one
outgoing electron from each Majorana bound state.

\section{Conclusions}

In this paper, we showed that a one dimensional device with two
Majorana bound states at its ends yields a Hanbury-Brown Twiss
effect in the internal electron-hole pseudospin space which may be
detected in realistic condensed matter settings through
current-current correlations. This is a departure from {\em all}
the known multi-particle interference effects which have
manifested themselves in spatial bunching and antibunching or
spin-spin correlations. Fundamentally, it can be regarded as a
manifestation of the quantum indistinguishability between
electronic annihilation and hole creation evidencing the presence
of Majorana bound states. The same settings may also be used to
engineer a non-local entangler of distant electronic spins, which
may enable circumventing the decoherence arising from the
transport needed to separate entangled particles.

{\em Acknowledgements}: We thank  C. Chamon, R. Egger, R. Jackiw,
M. Pepper, S. Y. Pi, G. W. Semenoff and S. Tewari for fruitful
discussions. SB (PS) thanks the University of Perugia (University
College London) for hospitality and partial support. SB thanks the
EPSRC, UK, the Royal Society and the Wolfson Foundation.


\begin{thebibliography}{10}

\bibitem{Hong-Ou-Mandel}C. K. Hong, Z. Y. Ou, and L. Mandel, Phys. Rev. Lett. {\bf
59}, 2044 (1987).
\bibitem{yamamoto} R. C. Liu {\em et. al.}, Nature {\bf 391},
263-265 (1998).
\bibitem{Aspect} T. Jeltes {\em et. al.}, Nature {\bf 445}, 402 (2007).
\bibitem{Shih-Alley} Y. H. Shih and C. O. Alley, Phys. Rev. Lett. {\bf 61}, 2921–2924 (1988).
\bibitem{Bose-Home} S. Bose and D. Home, Phys. Rev. Lett. {\bf 88}, 050401 (2002).
\bibitem{Fu-Kane2} L. Fu and C. L. Kane, Phys. Rev. Lett. {\bf 102}, 216403 (2009); A. R. Akhmerov, J. Nilsson and C. W. J. Beenakker, Phys. Rev. Lett. {\bf 102}, 216404 (2009)
\bibitem{Sodano} G. W. Semenoff and P. Sodano, J. Phys. B: At. Mol. Opt. Phys. {\bf 40}, 1479 (2007).
\bibitem{Beenakker} J. Nilsson, A.R. Akhmerov, and C.W.J. Beenakker, Phys. Rev. Lett. {\bf 101}, 120403 (2008).
\bibitem{Fu} L. Fu, Phys. Rev. Lett. {\bf 104}, 056402 (2010).
\bibitem{Kitaev2} A. Yu. Kitaev,  Phys.-Usp. {\bf 44}, 131 (2001).
\bibitem{Das-Sarma1} S. Tewari {\em et. al}, Phys. Rev. Lett. {\bf 98}, 010506 (2007).
\bibitem{Fu-Kane1}L. Fu and C. L. Kane, Phys. Rev. Lett. {\bf 100}, 0964407
(2008).
\bibitem{Refael2} V. Shivamoggi, G. Refael, and J. E. Moore,  Phys. Rev. B {\bf 82}, 041405 (2010).
\bibitem{Das-Sarma2} J. D. Sau {\em et. al}, arXiv:1006.2829v2 (2010); J. D. Sau {\em et. al}, Phys. Rev. Lett. {\bf 104}, 040502 (2010).
\bibitem{Refael1} Y. Oreg, G. Refael, F. von Oppen, arXiv:1003.1145v2 (2010).
\bibitem{JackiwPi} C. Chamon {\em et. al.}, Phys. Rev. B {\bf 81}, 224515 (2010).
\bibitem{Kitaev1} A. Yu. Kitaev Ann. Phys. (N.Y.) {\bf 303}, 1 (2003); C. Nayak {\em et. al}, Rev. Mod. Phys. {\bf 80}, 1083 (2008).
\bibitem{datta} S. Datta {\em Electronic Transport in Mesoscopic
Systems}, (Cambridge University Press, 1995),pp. 145-157; D.S.
Fisher and P. A. Lee, Phys. Rev. B {\bf 23}, 6851 (1981).
\bibitem{ghur} T. Ghur, A. M\"{u}ller-Groeling, H. Weidenmuller,
Phys. Rep. {\bf 299}, 189-425 (1998).
\bibitem{buttiker} M. B\"{u}ttiker, Phys. Rev. Lett. {\bf 65}, 2901 (1990).
\bibitem{loss-burkard} G. Burkard, D. Loss, E. V. Sukhorukov, Phys. Rev. B {\bf 61}, R16303 (2000).
\bibitem{Bolech-Demler} C. J. Bolech and E. Demler, Phys. Rev. Lett. {\bf 98}, 237002 (2007).
\bibitem{martin} O. Sauret, T. Martin, D. Feinberg, Phys. Rev. B {\bf 72}, 024544 (2005).

\bibitem{entconc} C. H. Bennett {\em et. al}, Phys. Rev. A {\bf 53}, 2046 (1996).
\bibitem{Pepper} M.D. Blumenthal {\em al.}, Nat. Phys. {\bf 3} 343 (2007); S. J. Wright {\em et. al.}, Phys. Rev. B {\bf 80}, 113303 (2009).




\end{thebibliography}
\end{document}